# Investigating Spatial Error Structures in Continuous Raster Data

Narumasa Tsutsumida[a*], Pedro Rodríguez-Veiga[b,c], Paul Harris[d], Heiko Balzter[b,c], Alexis Comber[e]

[a] Graduate School of Global Environmental Studies, Kyoto University, Kyoto, 606-8501, Japan

[b] Centre for Landscape and Climate Research, University of Leicester, Leicester, LE1 7RH, UK

[c] NERC National Centre for Earth Observation (NCEO), University of Leicester, Leicester, LE1 7RH, UK

[d] Sustainable Agriculture Sciences, Rothamsted Research, North Wyke, Okehampton, EX20 2SB, UK

[e] School of Geography, University of Leeds, Leeds LS2 9JT, UK

[*] Corresponding author.

E-mail address: naru@kais.kyoto-u.ac.jp

Address: Yoshidahonmachi, Sakyo, Kyoto, 606-8501, Japan.

## Abstract

The objective of this study is to investigate spatial structures of error in the assessment of continuous raster data. The use of conventional diagnostics of error often overlooks the possible spatial variation in error because such diagnostics report only average error or deviation between predicted and reference values. In this respect, this work uses a moving window (kernel) approach to generate geographically weighted (GW) versions of the mean signed deviation, the mean absolute error and the root mean squared error and to quantify their spatial variations. Such approach computes local error diagnostics from data weighted by its distance to the centre of a moving kernel and allows to map spatial surfaces of each type of error. In addition, a GW correlation analysis between predicted and reference values provides an alternative view of local error. These diagnostics are applied to two earth observation case studies. The results reveal important spatial structures of error and unusual clusters of error can be identified through Monte Carlo permutation tests. The first case study demonstrates the use of GW diagnostics to fractional impervious surface area datasets generated by four different models for the Jakarta metropolitan area, Indonesia. The GW diagnostics reveal where the models perform differently and similarly, and found areas of under-prediction in the urban core, with larger errors in peri-urban areas. The second case study uses the GW diagnostics to four remotely sensed aboveground biomass datasets for the Yucatan Peninsula, Mexico. The mapping of GW diagnostics provides a means to compare the accuracy of these four continuous raster datasets locally. The discussion considers the relative nature of diagnostics of error, determining moving window size and issues around the interpretation of different error diagnostic measures. Investigating spatial structures of error hidden in conventional diagnostics of error provides informative descriptions of error in continuous raster data.

## Keywords

Error distribution, Spatial accuracy, Local error diagnostics, Spatial heterogeneity

## 1. Introduction

All spatial data are subject to error. Remotely sensed (RS) imagery routinely contains sensor-related errors, atmospheric effects, and geometric errors. Environmental datasets that describe landscape features and properties from RS products (e.g. forest aboveground biomass, species distribution, and climate change scenarios) inherently contain prediction errors. Errors can manifest themselves as systematic deviations and/or noise which require careful assessment in order to avoid mis-interpretations of the data, to support reliable conclusions and to make informed decisions (Daly, 2006; Foody, 2002). Error assessments provide a guide to data quality and reliability (Foody, 2002) and can provide earth observation (EO) scientists with an understanding of the sources of error both in RS imagery and products (Liu et al., 2007; Stehman and Czaplewski, 1998). However, conventional summary measures of error do not take any spatial information (e.g. spatial heterogeneity) of error into account (Foody, 2005, 2002). Spatially explicit approach for the assessment is hence important.

In EO studies, spatial extensions of conventional diagnostics of error or accuracy have been demonstrated for categorical raster data, such as land cover classification data (Comber et al., 2017, 2012; Comber, 2013; Congalton, 1988; Foody, 2005). These approaches spatially extend the usual method of estimating and reporting accuracy through a confusion matrix, which is the cross-tabulation of predicted and reference classes to generate measures of user's and producer's accuracy that correspond to commission and omission errors, respectively, along with an overall accuracy (Congalton, 1991; Stehman and Czaplewski, 1998). Specifically, Comber (2013) demonstrated the use of a geographically weighted (GW) approach to generate spatial surfaces of these measures. The GW approach calculates a series of local diagnostics of accuracy, using data weighted by their distance to the centre of a moving window or kernel to explore spatial heterogeneity (Gollini et al., 2015). This has been used to compare global land cover datasets (Comber et al., 2013), to assess the consistency of such classification over time (Tsutsumida and Comber, 2015), and to construct hybrid global land cover datasets from multiple inputs (See et al., 2015). Comber et al. (2017) proposed GW confusion matrices for further generic applications. The GW framework itself (Fotheringham et al., 2002; Gollini et al., 2015; Lu et al., 2014) has been widely adopted across many scientific disciplines (e.g. Geography, Ecology, Health), where GW regression (Brunsdon et al., 1996) is the most popular GW model.

The developments of spatially explicit approaches for error assessment in continuous raster data in the EO domain have been limited. Comber et al. (2012) proposed a fuzzy GW difference analysis which estimates absolute deviations between the predicted and reference fuzzy membership, essentially applying a fuzzy generalization of the categorical accuracy measures. Khatami et al. (2017) proposed a spatial interpolation approach for soft classification maps in which a linear kernel function was applied to interpolate spatial deviations between predicted and reference proportions, with a focus on weight of spectral or class proportion as a soft classification measure. Willmott and Matsuura (2006) described maps of cross-validation error. Continuous raster data are commonly assessed using mean signed deviation (msd), mean absolute error (mae), root mean square error (rmse) and Pearson's correlation coefficient ($r$). Accurate predictions are reflected by msd, mae and rmse to be zero, coupled with $r$ to be one. Although these conventional diagnostics are useful in reporting error, each of them provides an overall, global or 'whole map' measure only. In this respect, Harris and Juggins (2011) demonstrated GW $r$ for assessing UK freshwater acidification prediction accuracy. Harris et al. (2013) demonstrated GW mae for UK freshwater acidification and London house price prediction accuracy, as separate case studies. Monteys et al. (2015) demonstrated GW $r$ for assessing water depth prediction accuracy in Irish coastal waters. These studies either directly extend GW summary statistics (e.g. GW averages, GW variances) as first proposed by Brunsdon et al. (2002), or directly use GW $r$ (Fotheringham et al., 2002), but in a model accuracy context. Further advances of GW summary statistics can be found in Harris and Brunsdon (2010) and Harris et al. (2014). However, the previous studies have only reported spatial error briefly as part of a suite of diagnostics. That is, spatial extensions of conventional diagnostics of error for continuous raster data have not been described in a comprehensive way, specifically in an EO context. Here we demonstrate the linked use of all four diagnostics, msd, mae, rmse and $r$, through their GW msd, GW mae, GW rmse and GW $r$ counterparts and advance them through the application of Monte Carlo permutation tests to identify unusual clusters of error applied to two EO case studies. The first case study evaluates datasets of the fractional impervious surface area (%ISA) with the aim of investigating spatial structures of error in multiple predictions by four different models. The second case study evaluate four different forest aboveground biomass (AGB) datasets in order to compare spatial structures of error in multiple independent datasets.

## 2. Case study data

### 2.1 Study 1

In order to explore how spatial structures of error can differ according to different models, four independent predictions of %ISA in the Jakarta Metropolitan Area (JMA), Indonesia, for 2012 were produced. The %ISA was inferred from the enhanced vegetation index (EVI) stored in moderate resolution imaging spectroradiometer (MODIS) MOD13Q1 product, which are 16-days composite RS imagery with a 231 m spatial resolution. Annual minimum, mean, maximum, and standard deviation of EVI were calculated on a pixel by pixel basis from the 24 images in 2012. These data were classified and assessed using training and reference (validation) samples collected at 984 randomly selected grid squares of the same size and at the same locations as the MODIS MOD13Q1 product. The %ISA was visually interpreted from fine resolution images in available Google Earth from the same year (Comber et al., 2016; Tsutsumida et al., 2016; Tsutsumida and Comber, 2015). When fine resolution images were not available at a sampling grid in 2012, %ISA were interpolated from images dated before and after the year 2012, only if the %ISA is stable over the period (in most cases, %ISA is zero). It is a reasonable approach because impervious surfaces do not change frequently. The reference values of %ISA were interpreted twice to minimize human error. The sample grids were randomly divided into training (*n* = 434) and reference data (*n* = 550) as shown in Figure 1.

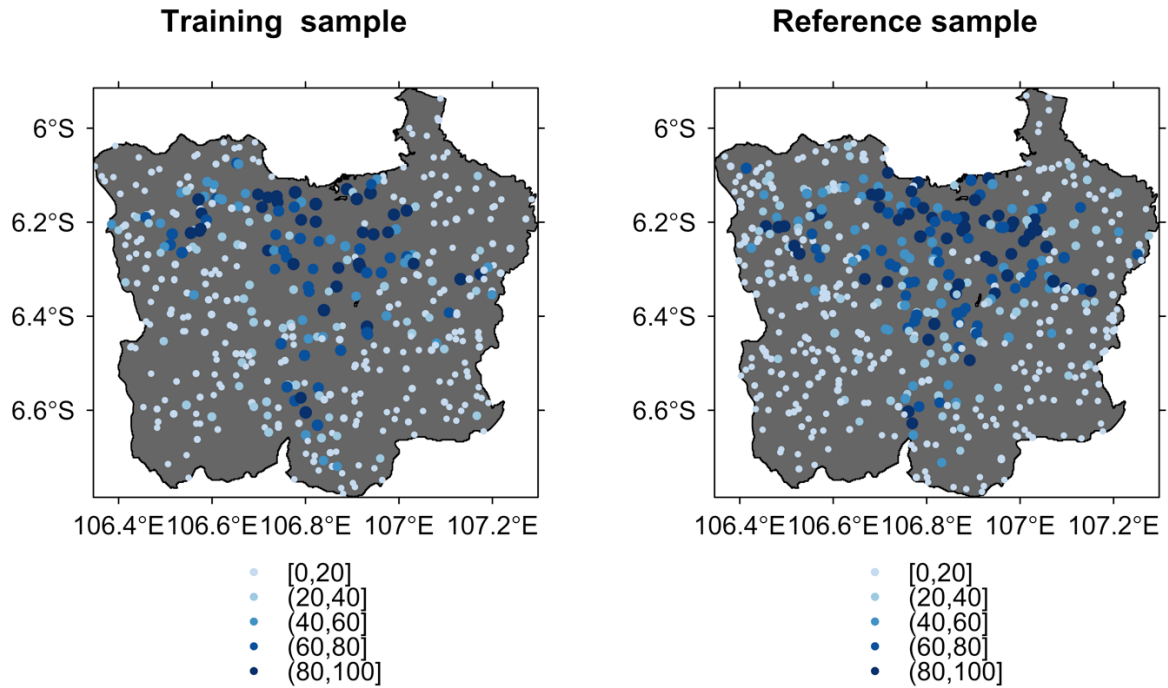

*Figure 1. The spatial distribution of the training (left) and reference (right) sample of fractional impervious surface area (%) in the Jakarta metropolitan area, Indonesia.*

Four different models were implemented to predict %ISA in the JMA: Logistic regression, Maximum Entropy (MaxEnt), Random Forest (RF) regression, and class probability of the RF classifier (hereafter RF probability). All four models return a continuous classification value between 0-100%. Logistic regression is a parametric generalized linear model for response data following a binomial distribution. The outcomes are within the range between 0 and 1 (rescaled to 0-100%). MaxEnt is a non-parametric model, which naturally extends from logistic regression (Phillips and Dudík, 2008). MaxEnt returns the probability of presence from presence-only training data (i.e. without labelled "absent" data), resulting %ISA predictions. RF regression and RF probability are machine learning techniques using ensemble logistic trees (Breiman, 2001). For RF regression, each tree is constructed by bootstrapped random sampling so that random sample selection leads to a weak correlation between trees. For RF probability, each tree votes for the most popular class and a random sample selection to grow trees is used to minimize the classification error. Due to its voting system, RF produces a probability of class presence, predicting %ISA. The %ISA predictions of these four models are different and clearly vary spatially (Figure 2). Note that apparent water surfaces are masked by a MODIS MOD44W product which represents the water surface in the same spatial scale of the MOD13Q1. Thus, submerged areas (e.g., those found in the North-East edge of the JMA) are excluded in this analysis.

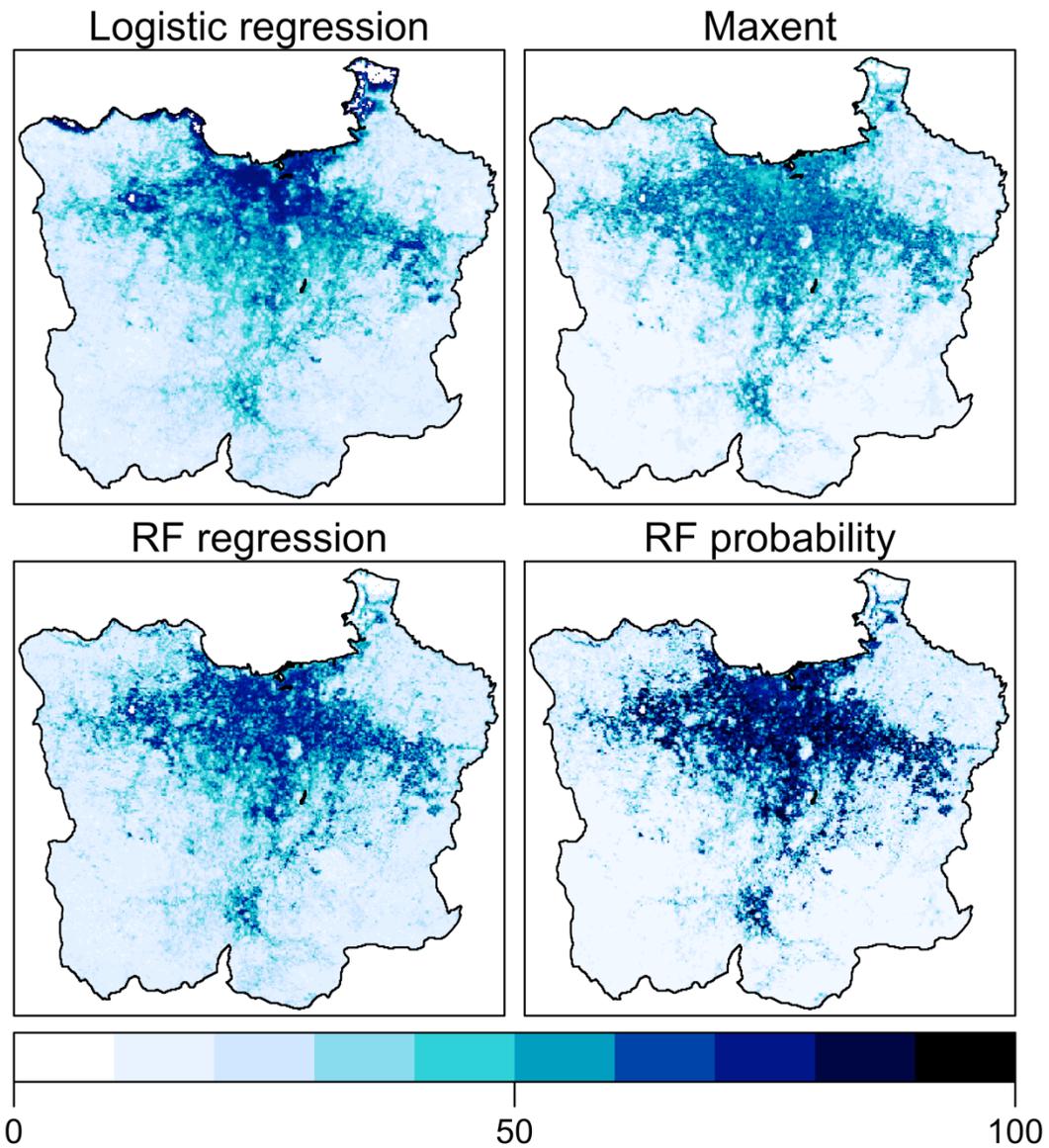

Figure 2. Predicted fractional impervious surface area (%) by four models for study 1: Logistic regression (Left upper), MaxEnt (Right upper), RF regression (Left bottom), and RF probability (Right bottom).

## 2.2 Study 2

In order to explore how spatial structures of error can differ according to available different datasets, four AGB spatial datasets for the Campeche, Yucatan, and Quintana Roo administrative regions in the Yucatan peninsula, Mexico are used (Figure 3). These were developed by Rodríguez-Veiga et al. (2016), Baccini et al. (2012), Saatchi et al. (2011), and Hu et al. (2016). Details of these datasets are summarized in Table 1. Dry forest, moist forest, and mangrove forest are found in the North-Western region, the central region, and the coastal zone of the Yucatan peninsula. It is not possible to objectively determine which dataset is the most accurate from Table 1, as the reported errors are derived from different reference sources. The reference data for this case study was provided by the INFyS *in-situ* observation data which record measures of AGB (Mg ha$^{-1}$) at four nested 0.04 ha subplots within 1 ha field plots (Rodríguez-Veiga et al., 2016). Data from a total of 286 (1 ha) field plots were used as reference measures of AGB for the period 2004-2007 (Figure 4). It is noted that the spatial resolution of assessed AGB datasets and reference sample is different, which is a limitation of data availability, similar to the study of Rodríguez-Veiga et al. (2016).

*Table 1. Descriptions of four forest aboveground biomass datasets used for study 2.*

| Dataset | Area | Spatial resolution | Period | Input and trained data | Method | Reported Accuracy |
|---|---|---|---|---|---|---|
| Rodríguez-Veiga et al. (2016) | Forest areas in the entire territory of Mexico | 250 m | 2008 | MODIS, Advanced Land Observing Satellite (ALOS) Phased Array type L-band Synthetic Aperture Radar (PALSAR) dual-polarization backscatter coefficient images, and the shuttle radar topography mission (SRTM) digital elevation model (DEM), trained with the INFys *in-situ* dataset | MaxEnt | Rmse of 36.1 Mg ha$^{-1}$ and $R^2$ of 0.31 |
| Baccini et al. (2012) | The pan-tropics forests in the world | 500 m | 2007-2008 | MODIS Nadir Bidirectional Reflectance Distribution Function Reflectance (BRDF), temperature, a DEM from the SRTM and Geoscience Laser Altimeter System (GLAS) data | RF | Rmse of 50 Mg ha$^{-1}$ for tropical America, 38 Mg ha$^{-1}$ for Africa, and 48 Mg ha$^{-1}$ for Asia |
| Saatchi et al. (2011) | The pan-tropics forests in the world | 0.0083 degree | The early 2000s | MODIS, SRTM, and quick scatterometer (QSCAT), as well as GLAS data input | Fusion model based on MaxEnt | Relative error of 27.3% for Latin America, 31.8% for Africa, and 33.4% for Asia |
| Hu et al. (2016) | Global | 1 km | 2004 | Multiple variables such as ground inventory data, optical imagery, GLAS, DEM, and climate data, incorporating over 4000 ground inventory observations from all over the world | RF | Rmse of 87.53 Mg ha$^{-1}$ |

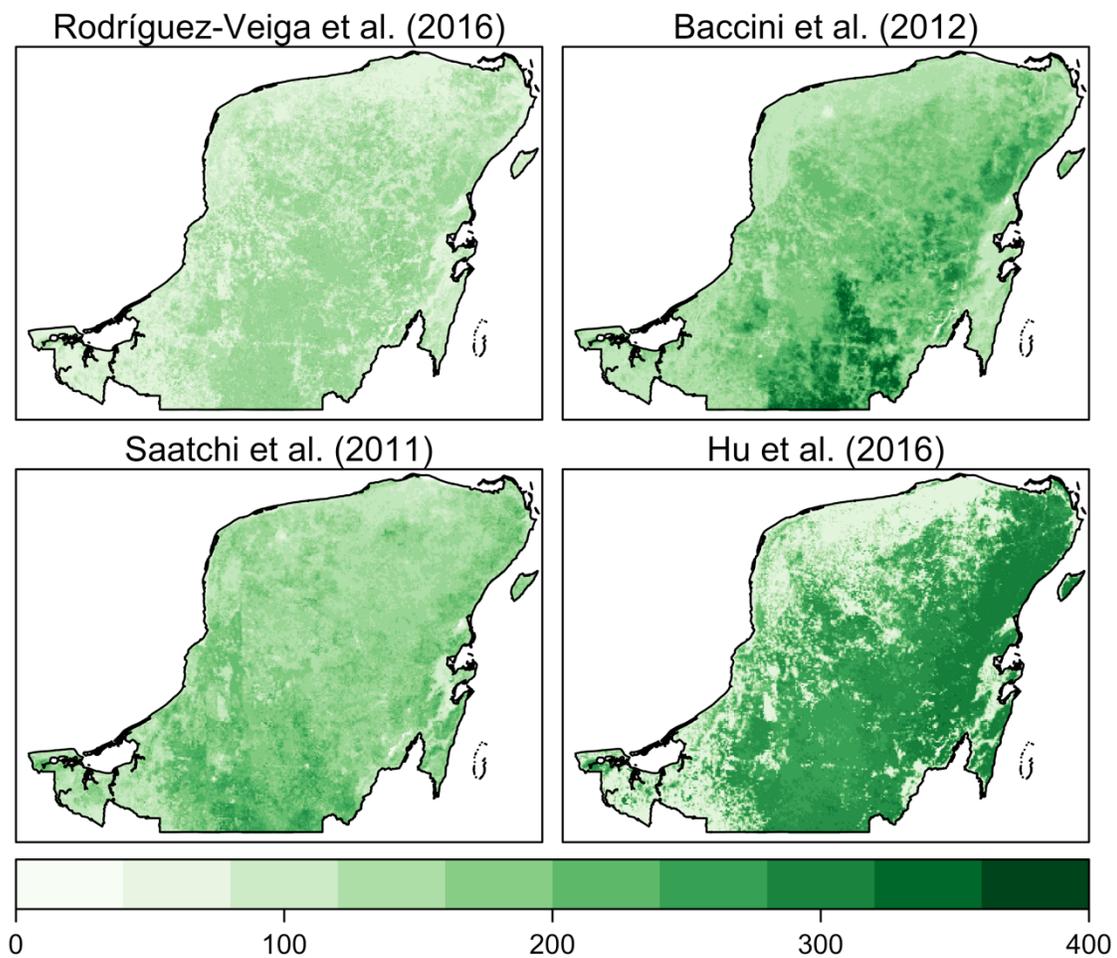

*Figure 3. Four aboveground biomass datasets (units: Mg ha$^{-1}$) for the Yucatan peninsula, Mexico for study 2.*

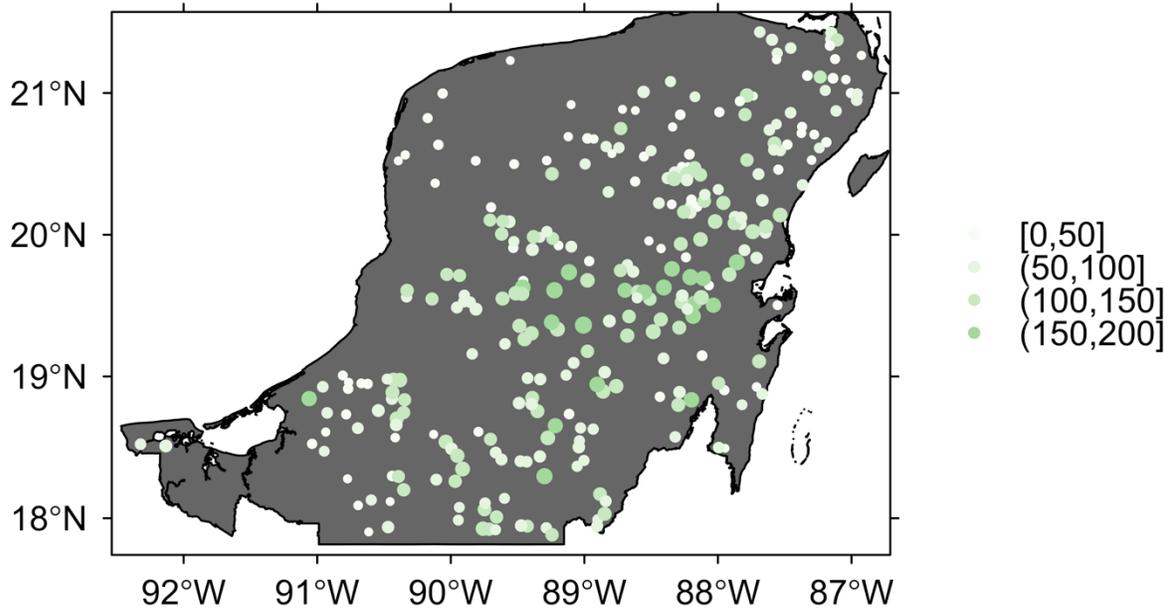

*Figure 4. The spatial distribution of in-situ reference sample points for forest aboveground biomass data (units: Mg ha$^{-1}$) in the Yucatan peninsula, Mexico.*

## 3. Methods

The GW versions of msd, mae, rmse, and *r* are described as follows. At any location $i$, GW msd: $gw.msd(x_i, y_i)$, GW mae: $gw.mae(x_i, y_i)$, and GW rmse: $gw.rmse(x_i, y_i)$ are defined as:

$$gw.msd(x_i, y_i) = \frac{\sum_{j=1}^{n} \omega_{ij}(y_j - x_j)}{\sum_{j=1}^{n} \omega_{ij}} \qquad (1)$$

$$gw.mae(x_i, y_i) = \frac{\sum_{j=1}^{n} \omega_{ij}|y_j - x_j|}{\sum_{j=1}^{n} \omega_{ij}} \qquad (2)$$

and

$$gw.rmse(x_i, y_i) = \sqrt{\frac{\sum_{j=1}^{n} \omega_{ij}(y_j - x_j)^2}{\sum_{j=1}^{n} \omega_{ij}}} \qquad (3)$$

, where $x_j$ and $y_j$ are the reference and predicted values at sample location $j$, respectively, $\omega_{ij}$ weights controlled by a distance-decay kernel function (Equation 8) with respect to location $i$ and $j$, and $n$ is the total number of sample data points. Observe that this always holds, msd ≤ mae ≤ rmse (Willmott and Matsuura, 2005) and their GW counterparts have the same characteristics.

A GW $r$ at any location $i$, is found using:

$$gw.cor(x_i, y_i) = \frac{c(x_i, y_i)}{s(x_i)s(y_i)} \qquad (4)$$

where a GW standard deviation: $s(x_i)$ is

$$s(x_i) = \sqrt{\frac{\sum_{j=1}^{n} \omega_{ij}(x_j - m(x_i))^2}{\sum_{j=1}^{n} \omega_{ij}}} \qquad (5)$$

and a GW mean: $m(x_i)$ is

$$m(x_i) = \frac{\sum_{j=1}^{n} \omega_{ij} x_j}{\sum_{j=1}^{n} \omega_{ij}} \qquad (6)$$

with a GW covariance: $c(x_i, y_i)$

$$c(x_i, y_i) = \frac{\sum_{j=1}^n \omega_{ij}\left[(x_j - m(x_i))(y_j - m(y_i))\right]}{\sum_{j=1}^n \omega_{ij}} \qquad (7)$$

For both case studies, the weights $\omega_{ij}$ are found using a bi-square kernel as follows:

$$\omega_{ij} = \begin{cases} \left(1 - \left(\frac{d_{ij}}{b}\right)^2\right)^2 & \text{if } |d_{ij}| < b, \\ 0 & \text{otherwise} \end{cases} \qquad (8)$$

, where $d_{ij}$ is the Euclidean distance between locations $i$ and $j$, and the kernel bandwidth $b$ is specified either as a fixed distance or an adaptive distance which includes a fixed number of data points for the local diagnostic calculation. In this study, an adaptive kernel was used as it suits the reference points of both case studies were not distributed uniformly. Its size was arbitrarily defined as 10% of nearby data to location $i$. The validity of this subjective bandwidth size is discussed in detail in section 5.

Observe that the chosen diagnostics complement each other: measures of msd, mae and rmse and their GW counterparts, all summarize the error in some manner, whilst $r$ and GW $r$ measure specifically the slope of the linear relationship between the predicted and reference values. Furthermore, $r$ and GW $r$ are scale invariant meaning that they cannot capture a consistent and uniform over- or under-prediction bias.

Fotheringham et al. (2002) presents methods for interpreting GW summary statistics (including GW $r$), and advocate Monte Carlo permutation tests. These tests can be adapted for GW error diagnostics (GW msd, GW mae, and GW rmse), in order to identify clusters where the diagnostics are 'significantly' or 'unusually' different to what would be found by chance or because of random variation in the error. Predicted and reference sample pairs are successively randomized (999 times in this study) and the local diagnostics are found after each randomization. A 'significance test' is then possible by comparing actual results with results from a large number of randomized distributions (i.e. by ranking all 1000 outcomes and ascertaining where the single, actual outcome lies). In this instance, the randomization hypothesis is that any pattern seen in the error occurs by chance and therefore any permutation of the error is equally likely. For GW $r$, the arguments are analogous, but where the investigation centers on the correlation between the predicted and reference values, rather than some summary of the error. In all instances, the permutation test should be viewed as informal and conditional on the GW diagnostic specification (i.e. bandwidth size, kernel type, etc.). Thus, throughout this study, the term 'significance' is used in an informal manner also, for this test.

In addition to calculating the global diagnostics of msd, mae, rmse and $r$, estimates and $p$-values for the significance of the Moran's $I$ of the deviation between predicted and reference values were calculated. These provide useful context and global information about spatial autocorrelation in the error. Weights were generated using an inverse distance squared function for the Moran's $I$ calculations.

## 4. Results

## 4.1 Study 1

Table 2 summarizes the conventional diagnostics of msd, mae, rmse, $r$ and the Moran's $I$ of %ISA predictions from logistic regression, MaxEnt, RF regression and RF probability. The negative msd values indicate that all four models under-predict, where RF regression provides the closest msd to zero (-2.95) and less errors than the other three models, with the smallest mae (15.51) and rmse (21.34) and the largest *r* (0.73). The logistic regression is the second most accurate with mae (15.77), rmse (21.87), and *r* (0.72). RF probability is the poorest predictor of %ISA as it shows the largest mae and rmse together with the smallest *r*, of all four models. All four models show significant spatial autocorrelations in their errors, where all *p*-values for the Moran's *I* estimates were less than 0.05. Nevertheless, no local spatial information about the errors is reported in Table 2.

*Table 2. Global diagnostics and Moran's I of fractional impervious surface area predicted by four different models for study 1.*

|  | msd | mae | rmse | *r* | Moran's *I*[*] |
|---|---|---|---|---|---|
| Logistic regression | -3.63 | 15.77 | 21.87 | 0.72 | 0.11 |
| MaxEnt | -7.57 | 15.83 | 22.74 | 0.72 | 0.11 |
| RF regression | -2.95 | 15.51 | 21.34 | 0.73 | 0.06 |
| RF probability | -5.12 | 15.85 | 24.22 | 0.71 | 0.05 |

[*] All *p*-values for estimates of Moran's *I* are less than 0.05.

Next, the spatial structure of the errors resulting from the %ISA predictions were explored using the three GW error diagnostics, together with GW *r* between the predicted and reference values as shown in Figure 5. Maps of GW msd indicate where the %ISA values are

over- or under-predicted, with positive values representing over-prediction. The GW mae and rmse maps reflect the magnitude of errors (absolute and root squared deviation, respectively, see also section 5). GW *r* depicts how the specified correlation varies across the JMA. Results for the associated Monte Carlo permutation tests are highlighted for *p*-values less than 0.01.

The GW msd results generally suggest that %ISA predictions are under-predicted when compared to reference values, especially in the urban core. Permutation tests locate 'significant' areas of unusually large, positive and negative GW msd values. A cluster of 'significantly' under-predicted values can be found in the middle of the JMA from all four models.

The GW mae and GW rmse maps show that peri-urban areas (surrounding the city core) tend to have larger mae/rmse values than others, suggesting the difficulty in predicting %ISA in complex urban frontiers between urban/non-urban areas. 'Significant' local clusters differ according to the models, but they tend to be distributed along such urban frontiers.

The results for GW *r* show South-Western and South-Eastern areas have consistently weak negative correlations in all four models, and the permutation tests indicate that such correlations are 'significantly' unusual for all models. As GW *r* represents spatial variation in the slope of the linear relationship between the predicted and reference values, maps of GW *r* can behave differently from those of the other three GW diagnostics which relate to error. Here only the GW mae and GW rmse maps show similarities to each other as expected (see section 5). As would also be expected from the results of Table 2, RF regression tends to provide the best local accuracy in most areas, but with clear spatial variation in this accuracy.

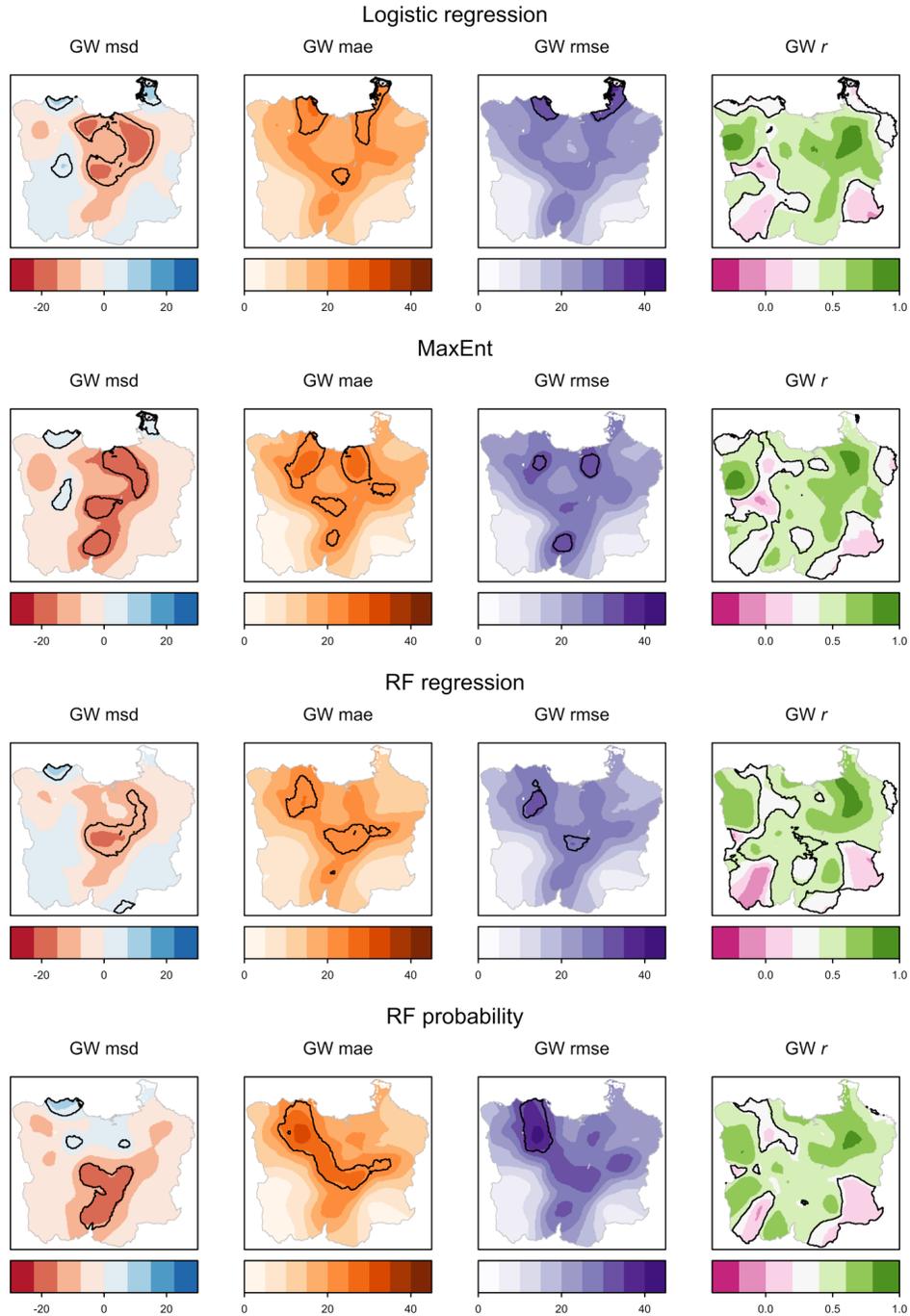

Figure 5. Spatial distributions of GW msd, GW mae, GW rmse, and GW r for fractional impervious surface area predicted from logistic regression, MaxEnt, RF regression, and class probability of RF classifiers for study 1. Black polygons represent 'significant' (p-values less than 0.01) areas by the Monte Carlo permutation tests.

## 4.2 Study 2

Conventional diagnostics for the four AGB datasets are shown in Table 3. Rodríguez-Veiga's dataset clearly provides the best accuracy amongst the four AGB datasets, as is evident from the closeness to zero of msd (-2.05), the smallest mae (31.52), the smallest rmse (39.42), and the largest $r$ (0.50). Hu's dataset is clearly the least accurate. The Moran's $I$ estimates are all significant ($p$-values were less than 0.05), indicating the possibility of existing a spatial structure to the errors in all four datasets.

*Table 3. Global diagnostics and Moran's I in forest aboveground biomass datasets for study 2.*

|  | msd | mae | rmse | $r$ | Moran's $I^*$ |
| --- | --- | --- | --- | --- | --- |
| Rodríguez-Veiga et al. | -2.05 | 31.52 | 39.42 | 0.50 | 0.08 |
| Baccini et al. | -86.36 | 89.10 | 105.58 | 0.36 | 0.35 |
| Saatchi et al. | -45.26 | 58.78 | 70.69 | 0.08 | 0.29 |
| Hu et al. | -136.35 | 141.76 | 153.75 | 0.36 | 0.24 |

$^*$ All $p$-values for estimates of Moran's $I$ are less than 0.05.

Figure 6 maps the three GW error diagnostics and GW $r$ in the four AGB datasets. Rodríguez-Veiga's dataset shows relatively small spatial variation in these diagnostics whilst Hu's dataset shows the largest variation. All four datasets perform very differently to each other with little spatial correspondence in their error.

In Rodríguez-Veiga's dataset, there is a 'significant' cluster of positive values of GW msd in the dry forests of the North-West, which is coupled with relatively small GW mae and GW rmse values and positive GW $r$ values. Forests in this area are often utilized for slash-and-burn agriculture, and the re-growth of trees can influence the remote sensing signals, resulting in potentially large prediction errors, but where it appears, not so large to adversely influence GW mae, and GW rmse, and GW $r$. Conversely, there are 'significant' clusters of negative values of GW msd in the moist forests of central-Eastern areas. These areas are coupled with 'significant' clusters of relatively large GW mae and GW rmse values and a 'significant' cluster of negative GW $r$ values. Thus, this dataset clearly performs worse in central-Eastern areas, as all four GW diagnostics indicate so. In this central-Eastern area, the forest is matured with large AGBs, so the saturation of spectral data from satellite sensors may be a cause of the inaccurate predictions.

Baccini's dataset depicts a 'significant' cluster of large positive GW msd values in the south, where the same area provides 'significantly' large GW mae, and GW rmse values, all suggesting an area of relatively poor AGB accuracy. Of note is the spatial behavior of GW $r$, where 'significant' negative correlations are of concern. Such clusters occur in quite different areas to the cluster observed in south for unusually large GW mae, and GW rmse values. Similar to the first case study, GW $r$ provides an alternative assessment of local error to GW mae and GW rmse. A possible explanation for this, is that GW $r$ can be sensitive to bandwidth size. For example, a few anomalous pairs of predicted and reference data points that fall close to the kernel centre can exert an undue influence on the correlation estimate (see section 5). In comparison to Rodríguez-Veiga's dataset, Baccini's dataset consistently performs worse in terms of AGB accuracy except a small portion of the central region in terms of GW mae.

Saatchi's dataset is relatively accurate in central areas with small GW msd, GW mae, and GW rmse values, whilst it is the least accurate the South-West, as confirmed by the permutation tests for GW msd, GW mae, and GW rmse, where 'significantly' large values are found. The GW *r* map shows negative values in many regions, but where no 'significant' clusters of this diagnostic are found. In comparison to Rodríguez-Veiga's dataset, Saatchi's dataset appears to perform better in some central-Eastern areas in terms of GW mae.

Hu's dataset depicts very different spatial patterns of the GW diagnostics to the other three datasets, and is clearly the least accurate with over-prediction almost everywhere. In particular, 'significantly' large GW msd, GW mae, and GW rmse values can be found in North-Eastern areas. A 'significantly' large negative GW *r* values are observed in the south but different areas from other three GW diagnostics.

In summary, mapping GW diagnostics provides useful spatial indications of the reliability of each dataset, not only individually, but also in comparison with each other. Despite all four datasets depicting the same AGB measure, the spatial patterns of error and accuracy vary in each dataset. Rodríguez-Veiga's dataset would be the best choice in terms of the conventional diagnostics (Table 3), but not necessarily the best choice everywhere, for example in central-Eastern areas, where Saatchii's dataset may be more accurate and preferred.

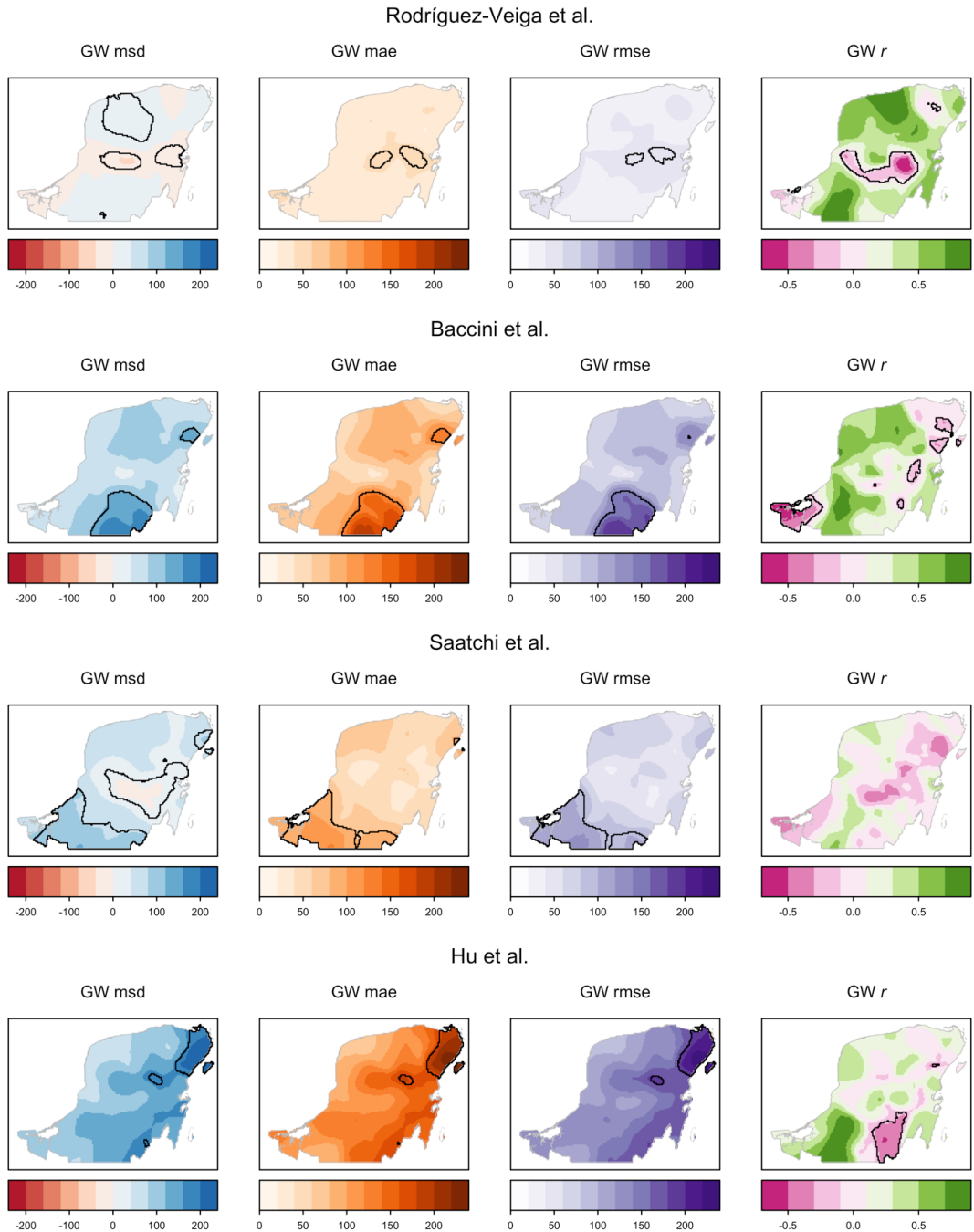

Figure 6. Spatial distributions of GW msd, GW mae, GW rmse, and GW r for four forest aboveground biomass datasets of Rodríguez-Veiga et al. (2016), Baccini et al. (2011), Saatchi et al. (2011), and Hu et al. (2016) for study 2. Black polygons represent 'significant' (p-values less than 0.01) areas by Monte Carlo permutation tests.

# 5. Discussion

The use of GW diagnostics has allowed investigations of the spatial structure of error between predicted and reference values for two EO case studies. This approach extends conventional (single-valued) whole map diagnostics of error spatially, through their localized (multiple-valued) counterparts. The associated permutation tests can highlight unusually accurate or unusually inaccurate error, providing a means to focus EO or other research activity on specific areas. This work is novel, but a number of points warrant discussion.

## 5.1 The effects of sample information

In this study, the use of the same reference data to evaluate the GW diagnostics of different datasets ensures results are comparable. However, an independent reference sample is not always available. This is a limitation for any error assessment: any results are only ever relative to the reference sample. For case study 2, Rodríguez-Veiga's dataset yielded the most accurate AGB performance amongst the four AGB datasets in most parts of the study area. The reference sample used here, despite being independent from the training data used for the Rodríguez-Veiga's dataset, originated from the same source (i.e. *in-situ* INFyS data), whilst the other three datasets used a completely different training dataset (i.e. GLAS footprints). Additionally, Rodríguez-Veiga's dataset is at a 250 m spatial resolution which is closer to the size of the reference data than the other datasets with spatial resolutions of 500 m or 1 km. Such characteristics need to be accounted for when comparing continuous raster datasets.

## 5.2 Bandwidth specification

In this work, a user-specified bandwidth of 10% was used for all outputs. This in part, reflected the need to use only one bandwidth throughout, so that multiple datasets could objectively be compared. However, in any GW approaches, the selection of the bandwidth is critical, to identify which levels of spatial heterogeneity should be focused. For example, Figure 7 shows the results of generating GW mae for Saatchi's dataset in case study 2, with a range of adaptive bandwidth sizes (5% to 50% in increments of 5%). Small bandwidth sizes result in highly localized variations in GW mae, while larger bandwidths result in a greater degree of smoothing and tend to be the global mae of 58.78 Mg ha$^{-1}$. Thus, results and interpretations are dependent on the user-specified 10% bandwidth. This can be overcome by calculating and visualizing a series of GW diagnostics over a range of bandwidths as an exploratory step. Although objective bandwidth selection procedures are available (Gollini et al., 2015; Harris et al., 2014), their use commonly results in one 'best on average' bandwidth choice that maximizes the precision of the predictor or statistic (e.g. via a leave-one-out cross-validation). Such data-driven procedures should not be regarded as a panacea for bandwidth selection or the degree of smoothing to use (Ruppert et al., 1995).

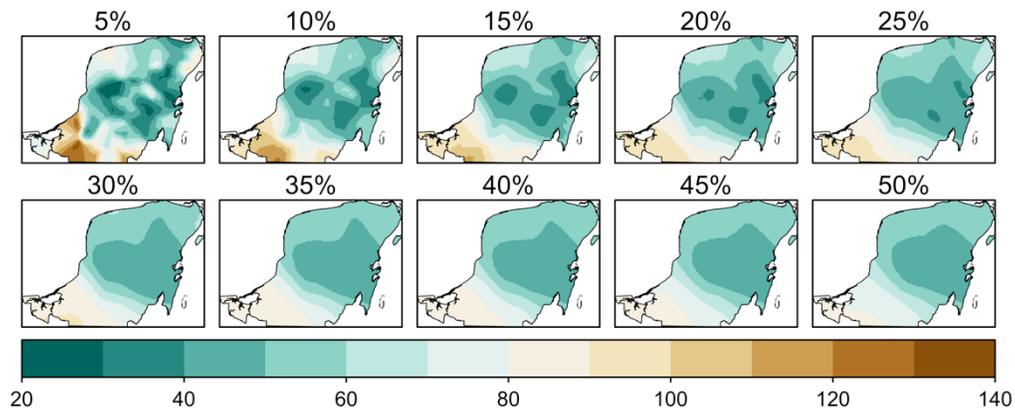

*Figure 7. Comparison of results of GW mae for Saatchi's aboveground biomass dataset with different adaptive kernel size between 5% to 50% as an example.*

## 5.3 The difference between mae and rmse

It is important to acknowledge the difference between mae and rmse, which is often overlooked. The mae represents average error magnitude (averaged absolute error) (Willmott and Matsuura, 2005), and rmse reflects the mean and variation in the error and is therefore highly sensitive to outliers (Pontius et al., 2008; Willmott and Matsuura, 2005). In this sense, mapping GW mae captures the spatial variation of the average error magnitude, whilst GW rmse highlights larger errors compared to GW mae. This is originated from the fact that the mean squared error, which is the squared rmse, is composed of the squared msd and the variance (Friedman, 1997). Because of this characteristic, rmse has no clear interpretation, unlike mae. A GW-based extension of such discussions would be an interesting topic of future work.

# 6. Conclusions

Conventional diagnostics of error, such as msd, mae and rmse provide global, 'on-average' measures. These summary measures of error do not capture any spatial information of the error. Ignoring spatial structures in error may result in a false interpretation and misuse of the data that the errors stem from. This work develops and applies localized diagnostics of error to investigate spatial heterogeneity of each types of these diagnostics. Two case studies demonstrate their value for comparing multiple models and for comparing multiple datasets. Comparing multiple models can support a deeper understanding of the spatial characteristics of errors and in turn can inform analytical choices and data collection to improve data accuracy and reliability. Identifying distinct local error clusters can help focus efforts in this respect. When multiple datasets are compared, understanding the spatial distributions of error in different datasets can inform choices about which datasets to use and in which areas to use them.

# Acknowledgements

We appreciate anonymous reviewers for giving insightful comments. This research is funded by KAKENHI Grant Number 15K21086; KU SPIRITS project; ROIS-DS-JOINT (006RP2018); and joint research program of CEReS, Chiba university (2018). P. Rodríguez Veiga and H. Balzter were supported by the UK's National Centre for Earth Observation (NCEO). A. Comber and P. Harris were supported by the Natural Environment Research Council Newton Fund grant (NE/N007433/1). We thank M. Castillo, I. Cruz, and M. Olguin for giving advice on the Mexican case study. All statistical analyses and mapping were conducted in the R open source software. Functions to calculate GW mae and GW rmse and their Monte Carlo permutation tests required a series of adaptions to the functions *gwss* and *gwss.montecarlo* in the GWmodel R package (Gollini et al., 2015; Lu et al., 2014). These adapted functions are available to interested researchers on request. Options for GW msd and GW *r* and their tests are already provided in the same functions of the GWmodel.

Harris, P., Clarke, A., Juggins, S., Brunsdon, C., Charlton, M., 2014. Geographically weighted methods and their use in network re-designs for environmental monitoring. Stoch. Environ. Res. Risk Assess. 28, 1869–1887. https://doi.org/10.1007/s00477-014-0851-1

Harris, P., Juggins, S., 2011. Estimating freshwater acidification critical load exceedance data for great britain using space-varying relationship models. Math. Geosci. 43, 265–292. https://doi.org/10.1007/s11004-011-9331-z

Hu, T., Su, Y., Xue, B., Liu, J., Zhao, X., Fang, J., Guo, Q., 2016. Mapping Global Forest Aboveground Biomass with Spaceborne LiDAR, Optical Imagery, and Forest Inventory Data. Remote Sens. 8, 565. https://doi.org/10.3390/rs8070565

Khatami, R., Mountrakis, G., Stehman, S.V., 2017. Predicting individual pixel error in remote sensing soft classification. Remote Sens. Environ. 199, 401–414. https://doi.org/10.1016/j.rse.2017.07.028

Liu, C., Frazier, P., Kumar, L., 2007. Comparative assessment of the measures of thematic classification accuracy. Remote Sens. Environ. 107, 606–616. https://doi.org/10.1016/j.rse.2006.10.010

Lu, B., Harris, P., Charlton, M., Brunsdon, C., 2014. The GWmodel R package: further topics for exploring spatial heterogeneity using geographically weighted models. Geo-spatial Inf. Sci. 17, 85–101. https://doi.org/10.1080/10095020.2014.917453

Monteys, X., Harris, P., Caloca, S., Cahalane, C., 2015. Spatial prediction of coastal bathymetry based on multispectral satellite imagery and multibeam data. Remote Sens. 7, 13782–13806. https://doi.org/10.3390/rs7101378232